\newfam\scrfam
\batchmode\font\tenscr=rsfs10 \errorstopmode
\ifx\tenscr\nullfont
        \message{rsfs script font not available. Replacing with
        calligraphic.}  \def\scr{\cal}
\else   
        \font\sevenscr=rsfs7 \font\fivescr=rsfs5 \skewchar\tenscr='177
        \skewchar\sevenscr='177 \skewchar\fivescr='177
        \textfont\scrfam=\tenscr \scriptfont\scrfam=\sevenscr
        \scriptscriptfont\scrfam=\fivescr \def\scr{\fam\scrfam}
        \def\cal{\scr}
\fi
\newfam\msbfam
\batchmode\font\twelvemsb=msbm10 scaled\magstep1 \errorstopmode
\ifx\twelvemsb\nullfont\def\Bbb{\bf}
         \font\eightbbb=cmb10 at 8pt
	\message{Blackboard bold not available. Replacing with
	boldface.}
\else   \catcode`\@=11
        \font\tenmsb=msbm10 \font\sevenmsb=msbm7 \font\fivemsb=msbm5
        \textfont\msbfam=\tenmsb \scriptfont\msbfam=\sevenmsb
        \scriptscriptfont\msbfam=\fivemsb
        \def\Bbb{\relax\expandafter\Bbb@} \def\Bbb@#1{{\Bbb@@{#1}}}
        \def\Bbb@@#1{\fam\msbfam\relax#1} \catcode`\@=\active
         \font\eightbbb=msbm8
\fi
        \font\eightrm=cmr8 \def\xrm{\eightrm} \font\eightbf=cmbx8
        \def\xbf{\eightbf} \font\eightit=cmti10 at 8pt
        \def\xit{\eightit}
        \font\eighttt=cmtt8 \def\xtt{\eighttt} \font\eightcp=cmcsc8
        \font\eighti=cmmi8 \def\xold{\eighti} \font\teni=cmmi10
        \def\old{\teni} \font\tencp=cmcsc10 \font\tentt=cmtt10
        \font\twelverm=cmr12 \font\twelvecp=cmcsc10 scaled\magstep1
        \font\fourteencp=cmcsc10 scaled\magstep2 
        \font\fourteenmath=cmmi12 scaled\magstep1
         \font\eightmath=cmmi8

\def\noblackbox{\overfullrule=0pt}
\noblackbox

\newtoks\headtext
\headline={\ifnum\pageno=1\hfill\else
	\ifodd\pageno{\eightcp\the\headtext}\dotfill{ }{\old\folio}
	\else{\old\folio}{ }\dotfill{\eightcp\the\headtext}\fi
\fi}
\def\makeheadline{\vbox to 0pt{\vss\noindent\the\headline\break
\hbox to\hsize{\hfill}}
        \vskip2\baselineskip}
\newcount\infootnote
\infootnote=0
\def\foot#1#2{\infootnote=1
\footnote{$\,{}^{#1}$}{\vtop{\baselineskip=.75\baselineskip
\advance\hsize by -\parindent\noindent{\xrm #2}}}\infootnote=0}
\newcount\refcount
\refcount=1
\newwrite\refwrite
\def\oldsize{\ifnum\infootnote=1\xold\else\old\fi}
\def\ref#1#2{
	\def#1{{{\oldsize\the\refcount}}\ifnum\the\refcount=1\immediate\openout\refwrite=\jobname.refs\fi\immediate\write\refwrite{\item{[{\xold\the\refcount}]}
#2\hfill\par\vskip-2pt}\xdef#1{{\oldsize\the\refcount}}\global\advance\refcount
by 1} }
\def\refout{\catcode`\@=11
        \xrm\immediate\closeout\refwrite \vskip2\baselineskip
        {\noindent\twelvecp References}\hfill\vskip\baselineskip
	\baselineskip=.75\baselineskip
        \input\jobname.refs 
	\baselineskip=4\baselineskip \divide\baselineskip by 3 
	\catcode`\@=\active\rm}

\def\hepth#1{\href{http://xxx.lanl.gov/abs/hep-th/#1}{{\xtt hep-th/#1}}}
\def\jhep#1#2#3{\href{http://jhep.sissa.it/stdsearch?paper=#1\%28#2\%29#3}{{\xrm JHEP} {\xbf #1} ({\xold#2}) {\xold#3}}}
\def\PLB#1#2#3{Phys. Lett. {\xbf B#1} ({\xold#2}) {\xold#3}}
\def\NPB#1#2#3{Nucl. Phys. {\xbf B#1} ({\xold#2}) {\xold#3}}
\def\PRD#1#2#3{Phys. Rev. {\xbf D#1} ({\xold#2}) {\xold#3}}

\newcount\sectioncount
\sectioncount=0
\def\section#1#2{\global\eqcount=0
	\global\subsectioncount=0 \global\advance\sectioncount by 1
        \xdef#1{{\old\the\sectioncount}} \vskip2\baselineskip\noindent
        \line{\twelvecp\the\sectioncount. #2\hfill}
        \vskip\baselineskip\noindent}
\newcount\subsectioncount
\def\subsection#1#2{\global\advance\subsectioncount by 1
	\xdef#1{{\old\the\sectioncount}.{\old\the\subsectioncount}}
	\vskip.8\baselineskip\noindent
	\line{\tencp\the\sectioncount.\the\subsectioncount. #2\hfill}
	\vskip.5\baselineskip\noindent}
\newcount\appendixcount
\appendixcount=0
\def\appendix#1{\global\eqcount=0
        \global\advance\appendixcount by 1
        \vskip2\baselineskip\noindent \ifnum\the\appendixcount=1
        \hbox{\twelvecp Appendix A:
        #1\hfill}\vskip\baselineskip\noindent\fi
        \ifnum\the\appendixcount=2 \hbox{\twelvecp Appendix B:
        #1\hfill}\vskip\baselineskip\noindent\fi
        \ifnum\the\appendixcount=3 \hbox{\twelvecp Appendix C:
        #1\hfill}\vskip\baselineskip\noindent\fi}

\newcount\eqcount
\eqcount=0
\def\Eqn#1{\global\advance\eqcount by 1
        \xdef#1{{\oldsize\the\sectioncount}.{\oldsize\the\eqcount}}
        \ifnum\the\appendixcount=0
        \eqno({\oldstyle\the\sectioncount}.{\oldstyle\the\eqcount})\fi
        \ifnum\the\appendixcount=1 \eqno({\oldstyle
        A}.{\oldstyle\the\eqcount})\fi \ifnum\the\appendixcount=2
        \eqno({\oldstyle B}.{\oldstyle\the\eqcount})\fi
        \ifnum\the\appendixcount=3 \eqno({\oldstyle
        C}.{\oldstyle\the\eqcount})\fi}
\def\eqn{\global\advance\eqcount by 1
        \ifnum\the\appendixcount=0
                \eqno({\oldstyle\the\sectioncount}.{\oldstyle\the\eqcount})\fi
                \ifnum\the\appendixcount=1 \eqno({\oldstyle
                A}.{\oldstyle\the\eqcount})\fi
                \ifnum\the\appendixcount=2 \eqno({\oldstyle
                B}.{\oldstyle\the\eqcount})\fi
                \ifnum\the\appendixcount=3 \eqno({\oldstyle
                C}.{\oldstyle\the\eqcount})\fi}
\def\multi{\global\advance\eqcount by 1}
\def\multieq#1#2{\xdef#1{{\oldsize\the\eqcount#2}}
        \eqno{({\oldstyle\the\eqcount#2})}}
\newtoks\url
\def\Href#1#2{\catcode`\#=12\url={#1}\catcode`\#=\active#2}
\def\href#1#2{{#2}}

\parskip=3.5pt plus .3pt minus .3pt
\baselineskip=14pt plus .3pt minus .05pt 
\lineskip=.5pt plus .05pt minus .05pt
\lineskiplimit=.5pt
\abovedisplayskip=18pt plus 4pt minus 2pt
\belowdisplayskip=\abovedisplayskip
\hsize=14cm
\vsize=20cm
\hoffset=1.5cm
\voffset=1.8cm
\frenchspacing
\nopagenumbers

\def\sss{\scriptscriptstyle}
\def\*{\partial}
\def\punkt{\,\,.}
\def\komma{\,\,,}

\def\+{\!+\!}
\def\={\!=\!}
\def\small#1{{\hbox{$#1$}}}
\def\half{\small{1\over2}}
\def\fraction#1{\small{1\over#1}}
\def\tr{\hbox{\rm tr}}
\def\eg{{\tenit e.g.}}
\def\ie{{\tenit i.e.}}

\def\nl{\hfill\break\indent}
\def\nlni{\hfill\break}

\def\\{\cr}
\def\*{\partial}

\def\d{\delta}

\def\D{\Delta}

\def\L{\Lambda}

\def\Z{{\Bbb Z}}
\def\R{{\Bbb R}}
\def\punkt{\,\,.}
\def\komma{\,\,,}

\def\II{\hbox{I\hskip-0.6pt I}}

\def\wdg{\!\wedge\!}
\def\fraction#1{\hbox{${1\over#1}$}}

\def\Fraction#1#2{\hbox{${#1\over#2}$}}
\def\tr{\hbox{\rm tr}}
\def\eg{{\tenit e.g.}}
\def\ie{{\tenit i.e.}}

\def\eq#1{$$#1\eqn$$}
\def\eql#1#2{$$#1\Eqn#2$$}
\def\eqa#1{$$\eqalign{#1}\eqn$$}
\def\eqal#1#2{$$\eqalign{#1}\Eqn#2$$}

\def\mx#1{\left(\matrix{#1}\right)}

\def\ra{\rightarrow}

\headtext={Cederwall, Gran, Nielsen, Nilsson: 
``({\eightmath p},{\eightmath q}) 5-Branes 
in Non-Zero {\eightmath B}-Field''}

%
%
%

\null\vskip-1cm
\hbox to\hsize{\hfill G\"oteborg-ITP-{\old99}-{\old18}}
\hbox to\hsize{\hfill\tt hep-th/9912106}
\hbox to\hsize{\hfill December, {\old1999}}

\vskip3cm
\centerline{\fourteencp ({\fourteenmath p},{\fourteenmath q}) 5-Branes
in Non-Zero {\fourteenmath B}-Field}
\vskip\parskip

\vskip1.2cm
\centerline{\twelverm Martin Cederwall, Ulf Gran, Mikkel Nielsen}
\centerline{\twelverm and Bengt E.W. Nilsson}

\vskip.8cm
\centerline{\it Institute for Theoretical Physics}
\centerline{\it G\"oteborg University and Chalmers University of 
Technology }
\centerline{\it SE-412 96 G\"oteborg, Sweden}

\vskip.8cm
\catcode`\@=11
\centerline{\tentt 
	martin.cederwall,gran,mikkel,bengt.nilsson@fy.chalmers.se}
\catcode`\@=\active

\vskip2.2cm

\centerline{\bf Abstract}

{\narrower\noindent We consider type \II B $(p,q)$ 5-branes in
constant non-zero background tensor potentials, or equivalently, with
finite constant field strength on the brane. 
At linear level, zero-modes are introduced and the physical degrees of 
freedom are found to be parametrised by a real 2- or 4-form 
field strength on the brane.  
An exact, SL(2;$\Z$)-covariant solution to the full non-linear 
supergravity equations is then constructed. The resulting 
metric space-times are analysed, with special emphasis on the limiting
cases with maximal values of the tensor.
The analysis provides an answer to how the various background tensor
fields are related to Born--Infeld degrees of freedom and to
non-commutativity parameters.
\smallskip}
\vfill\eject

\ref\Goldstone{T.~Adawi, M.~Cederwall, U.~Gran, B.E.W.~Nilsson 
	and B.~Razaznejad,\nl {\xit ``Goldstone tensor modes''},
	\jhep{9902}{1999}{001} [\hepth{9811145}].}
\ref\Finite{M.~Cederwall, U.~Gran, M.~Holm and B.E.W.~Nilsson,\nl
	{\xit ``Finite tensor deformations of supergravity
	solitons''}, \jhep{9902}{1999}{003} [\hepth{9812144}].}
\ref\DualActions{M.~Aganagic, J.~Park, C.~Popescu and J.H.~Schwarz,\nl
	{\xit ``Dual D-brane actions''}, \NPB{496}{1997}{215}
	[\hepth{9702133}]}
\ref\Rajaraman{R. Rajaraman, {\xit ``Solitons and instantons''},
	North-Holland, 1982;\nlni J.A. Harvey, {\xit ``Magnetic
	monopoles, duality and supersymmetry''}, \hepth{9603086}.}
\ref\KappaBranes{M.~Cederwall, A.~von~Gussich, B.E.W.~Nilsson 
	and A.~Westerberg,\nl {\xit ``The Dirichlet super-three-brane
	in type \II B supergravity''}, \nl \NPB{490}{1997}{163}
	[\hepth{9610148}];\nlni M. Cederwall, A. von Gussich,
	B.E.W. Nilsson, P. Sundell and A. Westerberg, \nl{\xit ``The
	Dirichlet super-p-branes in type \II A and \II B
	supergravity''}, \nl\NPB{490}{1997}{179}
	[\hepth{9611159}];\nlni M. Aganagic, C. Popescu and
	J.H. Schwarz, {\xit ``D-brane actions with local kappa
	symmetry''}, \nl\PLB{393}{1997}{311} [\hepth{9610249}];\nl
	{\xit ``Gauge-invariant and gauge-fixed D-brane actions''},
	\NPB{495}{1997}{99} [\hepth{9612080}];\nlni E. Bergshoeff and
	P.K. Townsend, {\xit ``Super D-branes''}, \NPB{490}{1997}{145}
	[\hepth{9611173}].}
\ref\CHSKM{C.G. Callan, Jr., J.A. Harvey and A. Strominger,
	{\xit ``Worldbrane actions for string solitons''},
	\nl\NPB{367}{1991}{60};\nlni D.M. Kaplan and J. Michelson,
	{\xit ``Zero modes for the D=11 membrane and five-brane''},
	\nl\PRD{53}{1996}{3474} [\hepth{9510053}].}
\ref\Duality{E.~Witten, 
	{\xit ``String theory dynamics in various dimensions''},
	\nl\NPB{443}{1995}{85} [\hepth{9503124}];\nlni C.M. Hull and
	P.K. Townsend, {\xit ``Unity of superstring dualities''},
	\nl\NPB{438}{1995}{109} [\hepth{9410167}];\nlni J.H. Schwarz,
	{\xit ``The power of M theory''}, \PLB{367}{1996}{97}
	[\hepth{9510086}];\nlni A.~Sen, {\xit ``Unification of string
	dualities''}, Nucl. Phys. Proc. Suppl. {\xbf58} ({\xold1997})
	{\xold5} [\hepth{9609176}];\nlni P.K. Townsend, {\xit ``Four
	lectures on M-theory''}, \hepth{9612121}.}
\ref\StringSolitons{M.J. Duff, R.R. Khuri and J.X. Lu,
	{\xit ``String solitons''}, Phys. Report. {\xbf259}
	({\xold1995}) {\xold213} [\hepth{9412184}].}
\ref\IIB{J.H. Schwarz,
        {\xit ``Covariant field equations of chiral N=2 D=10
        supergravity''},\nl \NPB{226}{1983}{269};\nlni
        P.S. Howe and P.C. West,
	{\xit ``The complete N=2, d=10 supergravity''},
	\NPB{238}{1984}{181}.}
\ref\JS{J.H. Schwarz, {\xit ``An SL(2;{\eightbbb Z}) multiplet of 
	type \II B superstrings''},
        \PLB{360}{1995}{13}\nl [\hepth{9508143}]; Erratum: ibid. {\xbf B364}
        ({\xold1995}) {\xold252}.}
\ref\DuffReview{M.J. Duff, {\xit ``Supermembranes''}, 
	\hepth{9611203}.}
\ref\LuRoy{J.X. Lu, S. Roy,
    {\xit ``On the construction of {\xrm SL(2,{\eightbbb Z})} type \II
	B 5-branes''}, \hepth{9812011}.}
\ref\CT{M. Cederwall and P.K. Townsend, {\xit ``The manifestly 
	{\xrm SL(2;{\eightbbb Z})}-covariant superstring''},
	\nl\jhep{9709}{1997}{003} [\hepth{9709002}]}
\ref\CW{M. Cederwall and A. Westerberg,
	{\xit ``World-volume fields, {\xrm SL(2;{\eightbbb Z})} and
		duality: the type \II B 3-brane''},
		\nl\jhep{9802}{1998}{004} [\hepth{9710007}].}
\ref\WeWy{A. Westerberg and N. Wyllard,
	{\xit ``Towards a manifestly {\xrm SL(2;{\eightbbb
	Z})}-covariant action for the\nl 
	type IIB (p,q) super-five-branes''}, \jhep{9906}{1999}{006}
	[\hepth{9905019}].}
\ref\SeibergWittenNC{N. Seiberg and E. Witten, 
	{\xit ``String theory and noncommutative geometry''},
	\nl\jhep{9909}{1999}{032} [\hepth{9908142}].}

\section\intro{Introduction}In investigating the structure of 
M/string theory, various extended non-perturbative objects,
``branes'', have proved to be very important [\Duality].  These
objects arise as solitonic solutions to the low-energy effective
theories [\StringSolitons,\DuffReview], \ie, supergravities in eleven
or ten dimensions. The physical degrees of freedom of the branes arise
as zero-modes around the solitonic solutions [\CHSKM,\Goldstone] in
analogy with the ordinary monopole case [\Rajaraman] (see however
ref. [\Goldstone] for some important differences due to whether the
theory contains gravity or not).  In a previous paper [\Goldstone], we
generalised the Goldstone prescription for generating zero-modes to
the case of tensor fields. This enabled us to treat all zero-modes on
an equal footing and also showed us exactly where the zero-modes sit
in the target space fields. In ref. [\Finite] we continued to explore 
the
zero-modes beyond linear level and were able to find exact solutions
to the full non-linear supergravity equations for the D3 and M5 branes
with finite constant field strength, or, equivalently, in the 
background of a constant
tensor potential. This paper is the natural
continuation of this work as we now turn to the more difficult case of
type \II B $(p,q)$ 5-branes.

In contrast to the D3 and M5 brane cases [\Finite], where we {\it a
priori} had strong reasons to expect analytic solutions, we now have
the additional complication of a real 4-form field strength on the
brane, which could make things worse [\DualActions]. In addition, the object used to
parametrise the zero-modes, a real 2- or 4-form, does not have any
duality property as in the previous D3 and M5 brane cases. This means
that the Ansatz for the finite solution will contain more terms and
that the algebra will be much more involved.

In section 2, we will review some properties of type \II B
supergravity that we will use and also write down the NS 5-brane
solution, which will be our starting point. The zero-mode solution,
which gives the parametrisation of the zero-modes, is obtained in
section 3. In section 4, we solve the full non-linear supergravity
equations and in section 5 we analyse the resulting metric
space-times.  We end by discussing the obtained solution and some
possible implications.

\section\prel{Preliminaries}The type \II B supergravity in ten 
dimensions 
has an SL(2;$\R$) invariance (which is broken to SL(2;$\Z$) 
by quantum effects) and contains two scalars in the coset space 
SL(2;$\R$)/U(1) (the dilaton $\phi$ and the axion $\chi$), a selfdual 
5-form field strength $H_{(5)}$, which is an SL(2;$\R$) singlet, and 
a 
real SL(2;$\R$) doublet of 3-form field strengths 
$H_{(3)r}=dC_{(2)r}$ ($r=1,2$), corresponding to the NS-NS and R-R
field strengths respectively. However, there exists a formulation,
which makes the SL(2;$\R$) covariance manifest [\IIB]. We will
use the notation of refs. [\CT,\CW]. The scalars are described by
a complex doublet,
${\cal U}^r$, which obey the following SL(2;$\R$)-invariant constraint
\eq{\Fraction{i}{2}\epsilon_{rs}{\cal U}^r\bar{{\cal U}}^s=1}
(we use conventions where $\epsilon^{12}=1$, $\epsilon_{12}=-1$).
Gauging the U(1) leaves two physical scalars which can be obtained through the projective invariant
$\tau={\cal U}^1/\,{\cal U}^2=\chi+i\, e^{-\phi}$. 
One advantage of the
covariant formulation is that the doublet of scalars transforms in a
simple way under SL(2;$\R$), whereas the physical scalars transform in
a more complicated way via the projective invariant $\tau$.  The
left-invariant SL(2;$\R$) Maurer--Cartan 1-forms are
\eq{
	Q=\fraction2\epsilon_{rs}d{\cal U}^r\bar{{\cal U}}^s\,
	,\quad\quad P=\fraction2\epsilon_{rs}d{\cal U}^r{\cal
	U}^s\punkt} 
Under the local U(1) transformation ${\cal
U}^r\ra{\cal U}^r e^{i\theta}$ (where the U(1) charge is
normalised to 1 for the scalar doublet), the 1-forms
transform as $Q\ra Q+d\theta\, ,\ P\ra P\,e^{2i\theta}$, \ie,
$Q$ is a U(1) gauge field and $P$ has U(1) charge 2. The
Maurer--Cartan equations are
\eq{
	DP=0\, ,\quad\quad dQ-i P\wdg\bar{P}=0\komma} 
where the
covariant derivative $D=d-ieQ$ acts from the right (although fermions are
not treated in this paper, we stick to standard superspace conventions), 
and $e$ is the
U(1) charge. The contravariant doublet of scalars can be combined
with a covariant doublet to yield an SL(2;$\R$)-invariant object. The
doublet can of course be retrieved from the scalar doublet and the
SL(2;$\R$)-invariant object, \eg,
\eq{{\cal H}_{(3)}\equiv{\cal U}^r H_{(3)r}\,,\qquad 
H_{(3)r}=\epsilon_{rs}\,{\rm Im}({\cal U}^s\bar{{\cal 
H}}_{(3)})\punkt} 
Using this formalism, the equations of motion can be written as
\eqal{
	&D{\ast}P+i\,{\cal H}_{(3)}\wdg\ast{\cal H}_{(3)}=0\komma\cr
	&D{\ast}{\cal H}_{(3)}-i\ast\bar{\cal
	H}_{(3)}\wdg P-4\,i\,H_{(5)}\wdg{\cal H}_{(3)}=0\komma\cr}{\eqms}
and the Bianchi identities are
\eqal{
	&D{\cal H}_{(3)}+i\,\bar{\cal H}_{(3)}\wdg P=0\komma\cr
	&dH_{(5)}-\Fraction{i}{2}{\cal H}_{(3)}\wdg\bar{\cal
	H}_{(3)}=0\punkt\cr}{\bis} We also have the Einstein equations
\eq{R_{MN}=2 \bar P_{(M} P_{N)}+\bar{\cal H}_{(M}{}^{RS}{\cal 
H}_{N)RS}-\fraction{12}g_{MN}\bar{\cal H}_{RST}{\cal 
H}^{RST}+\fraction{6}H_{(M}{}^{RSTU}H_{N)RSTU}\punkt}

In type \II B we have an NS 5-brane, which couples magnetically to
$C_{(2)1}$ and a D5-brane, which couples magnetically to $C_{(2)2}$
(these are of course not the only tensor couplings---any $(p,q)$
5-brane couples to all tensors in type \II B supergravity, as can be
seen from the known D-brane actions [\KappaBranes]). Here we conventionally denote the charge doublet $(p,q)$,
in the covariant formalism this becomes $p_r$ ($r$=1,2), and for a
classical 5-brane these are just real numbers. Due to the existence of
$(p,q)$ strings [\JS], the charges will be quantised and for the
quantum mechanically allowed 5-branes, $p_r$ will be integers. The NS
5-brane corresponds to the charges (1,0) and the D5-brane to
(0,1). The physical charges are obtained by multiplying with the 
5-brane
charge quantum $\mu$.  Apart from the charge, the NS 5-brane solution 
is
characterised by the asymptotic value $\tau_\infty$. In the complex
formalism, the solution for asymptotically vanishing dilaton and axion
(\ie, $\tau_\infty=i$) is
\eqal{
	&ds^2=\Delta^{-1/4}dx^2+\Delta^{3/4}dy^2\komma\cr &{\cal
	H}^0_{(3)}=-\Fraction{i}{2}\,\Delta^{-1}\ast_y
	d\Delta\komma\qquad\qquad H_{(5)}=0\komma\cr
	&P=\Fraction{i}{4}\,\Delta^{-1}d\Delta\komma\qquad\qquad\qquad\quad
	Q=0\komma\cr &{\cal
	U}^1=i\,\Delta^{-1/4}\komma\qquad\qquad\qquad\quad{\cal
	U}^2=\Delta^{1/4}\komma\cr}{\bgsolution} 
where
$\Delta=1+\mu/\rho^2$, $\rho$ being the radial "distance" to the
brane, and $\ast_{x,y}$ means dualisation with respect to the
longitudinal ($x$) or the transverse ($y$) space. One
advantage of the complex formalism is that the above
expressions for the metric, the 3-form and the 1-forms contain
all the $(p,q)$ 5-branes.
The point is that we have doublets ${\cal U}^r$ and $H_{(3)r}$ which
transform under SL(2;$\R$) as ${\cal U}^r\rightarrow M^r{}_s{\cal U}^s$, 
$H_{(3)r}\rightarrow (M^{-1})^s{}_r H_{(3)s}$,
where
$M$ is an SL(2;$\R$) matrix. When contracting the indices, we
obtain an SL(2;$\R$)-invariant object. Hence the 3-form ${\cal
H}_{(3)}$ is the same for the NS 5-brane and the $(p,q)$
5-branes, as long as the background scalars are transformed
along with the charges. If we want a specific type of brane,
we can start with the NS 5-brane and then simultaneously make
an SL(2;$\Z$) transformation on the scalar and 3-form
doublets.

The general version of eq. (\bgsolution) for arbitrary charge $(p,q)$
and {\it arbitrary} background scalars is formally identical to
(\bgsolution) except for
\eql{{\cal U}^r=\epsilon^{rs}\bigl(-k\Delta^{1/4}p_s
+ik^{-1}\Delta^{-1/4}\tilde p_s\bigr)\komma}{\genbgsolution}
where $\epsilon^{rs}p_r\tilde p_s=1$ ($\tilde p_r$ does not have to be
integer) and $k$ is real, and where $\Delta=1+k^{-1}\mu/\rho^2$.  
Notice that $k^{-1}=|{\cal U}^r_\infty p_r|$, so the expression for 
$\Delta$
agrees with the one found in ref. [\LuRoy].
The asymptotic value of the physical scalar is
$\tau_\infty=-(kq-ik^{-1}\tilde q)/(kp-ik^{-1}\tilde p)$. 
Most of the
explicit calculations will, for the sake of economy of notation, 
make use of the NS 5-brane charge
$(p,q)=(1,0)$ and $\tau_\infty=i$, although the results will
be stated in SL(2;$\Z$)-covariant form.

\section\zmsol{Zero-mode Solution}We start by analysing the 
zero-modes 
using the general Goldstone
prescription introduced in ref. [\Goldstone]. This analysis will tell 
us
how the zero-modes are parametrised, in our case by a real 2- or
4-form in the longitudinal directions, and enable us to write down an
Ansatz for the exact solution. From the zero-mode solution we will
also see exactly where the zero-modes sit in the target space fields.

As described in \eg\ ref. [\Rajaraman], physical modes correspond to 
broken
{\it large} gauge transformations. By a large gauge transformation we
mean a transformation that does not go to zero at spatial infinity and
therefore will change the charges associated with an object. If \eg\ a
large gauge transformation changes the momentum of an object we must
conclude that the transformation corresponds to a time-dependent
translation. In this sense, large gauge transformations are equivalent
to global symmetries of the theory. By taking the equations of motion 
into
account, as we will see later, we can determine the transversal
dependence of the zero-modes and thus single out a particular large
gauge transformation which corresponds to a global symmetry. We then
have the ordinary correspondence between broken global symmetries and
physical modes.

{\it Small} gauge transformations, \ie , transformations that go to
zero at spatial infinity, on the other hand, relate equivalent
configurations and therefore have nothing to do with global symmetries
or zero-modes. They only represent a redundancy in our description of
the theory.

It is important to note that since we are considering a theory with
gravity,
\ie , with reparametrisation
invariance, in contrast to the ordinary monopole case, we must be
careful when we talk about \eg\ translations. To actually change the
brane configuration we must do something that is more than a small
gauge transformation and therefore have a non-vanishing effect at
infinity. What we really mean by a ``translation'' in a theory with
gravity is a {\it large} reparametrisation, and instead of saying that
broken translational invariance gives rise to scalar modes on the
brane we should say that the scalar modes arise by breaking the {\it
large} reparametrisation invariance in the transverse directions
[\Goldstone].  Since we actually consider a theory with {\it
super}gravity, the same arguing applies for the supersymmetry
transformations [\Goldstone].

We now introduce the tensor zero-modes as described in ref. 
[\Goldstone],
by making a gauge transformation $\d C_{(2)r}=d\L_{(1)r}$.
Using the proposed mechanism to generate zero-modes as large gauge
transformations, we make an Ansatz ${\cal U}^r\L_{(1)r}={\rm Re}{\cal
A}_{(1)}\D^{k_R}+i\, {\rm Im}{\cal A}_{(1)}\D^{k_I}$, where ${\cal
A}_{(1)}={\cal U}^r A_{(1)r}$ is a complex 1-form potential which lies
in the longitudinal directions and $A_{(1)r}$ is {\it constant}. That
the correct thing to do is to allow different radial behaviour for the
real and imaginary parts is understood from the background values of
${\cal U}^r$. The reason why we take ${\cal A}_{(1)}$ to lie in the
longitudinal directions is of course that we want to be able to
integrate out the transversal dependence, thus obtaining an effective
theory on the brane world-volume. We get
\eql{\d {\cal C}_{(2)}={\rm Re}{\cal A}_{(1)}\wdg d\D^{k_R}+i\, 
{\rm Im}{\cal A}_{(1)}\wdg d\D^{k_I}\punkt
}{\deltac} 
We now let $A_{(1)r}$ become $x$-dependent, which means
that (\deltac) is no longer a pure gauge transformation. By computing
$\d {\cal H}_{(3)}$ and solving the equations of motion for the
variation we will get the equations of motion and the transversal
behaviour of the zero-modes. We find (this relies on the observation that
the scalars do not transform at linear level---this may be deduced 
group-theoretically or {\it a posteriori} from the solution)
\eq{\d {\cal H}_{(3)}={\cal U}^r d(\d C_{(2)r})=
-{\rm Re}{\cal F}_{(2)}\wdg d\D^{k_R}-i
\,{\rm Im}{\cal
F}_{(2)}\wdg d\D^{k_I}\komma } 
where ${\cal F}_{(2)}={\cal U}^rF_{(2)r}={\cal U}^r dA_{(1)r}$ 
is the complex field strength on the
brane. (Since type \II B theory contains a doublet of 2-form potentials any
SL(2;$\Z$)-covariant description of a brane must contain a doublet 
of vector potentials. How the
complex field strength is related to the Born-Infeld field strength will be
clarified below.)
When solving the equations of motion it is convenient to have
all the $\D$-dependence explicit and we therefore introduce
${\cal\underline F}_{(2)}$, which is a $\D$-independent complex 2-form
when expressed in Lorentz indices, \ie, when we use the vielbeins to
go from coordinate-frame to Lorentz indices.  In the next section we
need to introduce higher matrix powers of ${\cal\underline
F}_{(2)}$. Using Lorentz indices, we then avoid the complexity of
metric factors.
     
We then have
\eq{\d {\cal H}_{(3)}=-{\rm Re}{\cal\underline F}_{(2)}\wdg 
d\D^{k_R}-i\, 
{\rm Im}{\cal\underline F}_{(2)}\wdg d\D^{k_I}\D^{-1/2}\komma}
where we have taken the $\D$ factors coming from the vielbeins and
from the $\cal U$'s into account.

We also have to introduce zero-modes in $H_{(5)}$. This will be clear 
by
looking at the result, but can be understood by the fact that not only
strings but also 3-branes may end on 5-branes, the two types of 
couplings
being related by what will manifest itself as a duality relation. 
Here, a slight complication is present. The Bianchi identity for the
5-form is readily solved by
$$
H_{(5)}=dC_{(4)}+\half\hbox{Im}({\cal C}_{(2)}\wdg\bar{\cal H}_{(3)})
\komma\Eqn\HfiveBIsol
$$
which is invariant under 
$$
\delta C_{(4)}=d\Lambda_{(3)}
+\half\hbox{Im}(\Lambda_{(1)r}{\cal U}^r\wdg\bar{\cal H}_{(3)})
\punkt\Eqn\deltaCfour
$$
The variation of $C_{(2)r}$ introduced in eq. (\deltac) clearly 
affects
$H_{(5)}$, in such a way, however, that the new 5-form field strength
is not single-valued on the transverse 3-sphere (remember that the 
background value of $C_{(2)r}$ is such that $H_{(3)r}$ wrap the sphere
with charges $(p,q)$). In order to use eq. (\HfiveBIsol) in a 
consistent
way, \ie, in order to work with transformations at the level of 
gauge potentials, two things have to be achieved by the 3-form gauge 
parameter
$\Lambda_{(3)}$. It must generate a ``large'' gauge transformation,
{\it and} it must, when made $x$-dependent, cancel the otherwise
inconsistent non-uniqueness in $H_{(5)}$. To be precise, the term we 
must
include in $\Lambda_{(3)}$, in addition to the zero-modes, is
\eq{\half{\rm Im}({\cal C}_{(2)}\wdg\Lambda_{(1)r}{\cal U}^r)\komma
}
where ${\cal C}_{(2)}$ is the background value of the potential and 
the
$x$-dependence of $\Lambda_{(1)r}$ is, as usual, turned on when the
variation of $C_{(4)}$ is obtained. In the present case, where
the background field strengths are purely transversal, the effect of
this is that the ``modification terms'' in $H_{(5)}$ as well as in
$\delta C_{(4)}$ are discarded, and one may formally proceed as if
``$H_{(5)}=dC_{(4)}$''. Would there have been longitudinal components
of background field strengths, one had been led to consider 
modified field strengths also in brane directions, as in ref. [\CW].

We thus make a gauge variation $\d C_{(4)}=d\L_{(3)}$, 
where $H_{(5)}=dC_{(4)}$. Using the Ansatz
$\L_{(3)}=\D^k A_{(3)}$, where $A_{(3)}$ is a constant 3-form
potential in the longitudinal directions, yields
\eq{\d C_{(4)}=-d\D^k\wdg A_{(3)}\punkt
} By letting $A_{(3)}$ become $x$-dependent we obtain
\eq{\d H_{(5)}=-d\D^k\wdg G_{(4)}-\ast_y d\D^k\wdg \ast_x 
G_{(4)}\komma
} where $G_{(4)}=dA_{(3)}$ and we have taken the requirement that
$H_{(5)}$ must be self-dual into account. To make the $\D$-dependence
explicit we introduce ${\underline G}_{(4)}$, which is
$\D$-independent when expressed in Lorentz indices,
\eq{\d H_{(5)}=-d\D^k \D^{-1/2}\wdg{\underline G}_{(4)}
-{\underline\ast}_y d\D^k \D^{1/2}
\wdg{\underline\ast}_x {\underline G}_{(4)}\komma} 
where $\underline\ast$ means dualisation with respect to the flat metric.

By inserting the variations into the second Bianchi identity in (\bis)
we get
\eq{d\ast_x {\underline G}_{(4)}=0\komma
} which is the equation of motion for ${\underline G}_{(4)}$. We also
get a relation between ${\underline G}_{(4)}$ and ${\cal\underline
F}_{(2)}$,
\eq{\ast_x {\underline 
G}_{(4)}=		
{k_R\over 2k(k_R-\fraction{4})}\,{\rm Re} 
{\cal\underline F}_{(2)}\komma} 
and the requirement that $k=k_R+\fraction{4}$.  The LHS of the first
identity in (\bis) vanishes to first order in ${\cal\underline
F}_{(2)}$.  Inserting the variations into the second equation in
(\eqms) yields the equation of motion for ${\cal\underline F}_{(2)}$,
\eq{d\ast_x {\cal\underline F}_{(2)}=0\komma
} and that $k_R=-\Fraction{3}{4}$ or $k_R=\Fraction{5}{4}$ and $k_I=0$
or $k_I=\Fraction{3}{4}$.  The LHS of the first equation in (\eqms)
vanishes to first order in ${\cal\underline F}_{(2)}$.

Since a BPS brane in $D=10$ has eight fermionic degrees of freedom,
we must have eight bosonic degrees of freedom in order to have
supersymmetry. The scalar fields contribute with four bosonic
degrees of freedom for a 5-brane and we must get the remaining four 
from
the tensor zero-modes. At first, it seems like we get too many
from the tensor modes, since a complex 2-form 
field strength in six dimensions has eight degrees of freedom, 
\ie, twice the amount we wanted. We,
however, require the zero-modes to be normalisable, which singles out
$k_R=-\Fraction{3}{4}$ and $k_I=0$, \ie, the imaginary part of $\d
{\cal H}_{(3)}$ vanishes. We have thus found that the zero-modes are
parametrised by a real 2-form field strength, ${\rm Re}{\cal\underline
F}_{(2)}$, which gives us the correct number of degrees of freedom.

We would like to comment in a more detailed way on the relation 
between
the field strength ``on the brane'' and the background  tensor 
potential. For an infinitesimally thin brane embedded in a target 
space, the equivalence is a consequence of the gauge invariance of
``$F-B=dA-B$''. In the context of a soliton solution, the brane is the
entire field configuration between the horizon and Minkowski infinity,
and the background fields are the values of the fields at infinity.
Here we do not have $F$ and $B$ as independent objects, rather the
behaviour of $B$ is parametrised in terms of $F$.
Take \eg\ the linearised solution of the present section. For a 
constant Re${\cal\underline F}_{(2)}$ we may write
the (interesting part of the) 3-form field strength as
$\delta{\cal H}_{(3)}=d\,[-\hbox{Re}{\cal\underline F}_{(2)}\Delta^{-3/4}]$.
Note that the potential inside the square brackets goes to zero when
$\rho$ goes to zero (at the horizon) and to a finite background value 
when 
$\rho$ goes to infinity. The configuration is that of an NS 5-brane
in a constant (longitudinal) 2-form background. This was demonstrated
here at linear level, but of course continues to hold to all orders for
the solutions presented in the following section.

In a type \II B theory, we are generically in a situation where there
are several tensor potentials coupling to a brane. An important piece
of information is the identification of the combination of tensor
potentials that are related to the Born--Infeld degrees of freedom,
and to the non-commutativity parameter. The parametrisation of the
zero-modes, and later of the full solutions, in terms of a real 2-form
provides exactly this information.

\section\finsol{Finite Field Strength Solution}The excitation of the 
3-form 
is parametrised by the $\Delta$-independent 2-form 
$F\equiv {\rm Re}\underline{\cal F}_{(2)}$, introduced in the 
previous section. The Goldstone analysis yields zero-modes which are 
linear in $F$, \ie, the analysis is only valid for small field 
strength. 
Finite constant field strength can be obtained by doing an 
expansion in $F$. However, we can truncate the expansion because of 
the 
Caley--Hamilton relation, which for antisymmetric matrices in 
6-dimensional Minkowski space takes the form
\eq{F^6=\fraction{2}\,t_2 
F^4+(\fraction{4}\,t_4-\fraction{8}\,t_2^2)\,
F^2+{\rm det}(F)\, \eta\,,}
where $t_2=\tr(F^2)$, $t_4=\tr(F^4)$ and $\eta$ is the diagonal matrix
corresponding to the Minkowski metric $(-,+,\ldots,+)$.

For the sake of simplicity, we start out with an NS 5-brane in 
$\tau_\infty=i$. The general case is kept under control as in section
\prel, and the SL(2;$\Z$)-covariant result will be stated at the end.
The charges are determined by the real doublet of
3-forms $H_{(3)r}$. Once we excite the 3-form we must also excite the
1-form $P$ and the 5-form to get solutions to the equations of
motion. Since $P$ is excited, the scalars will change. To stay on the
(1,0) brane, we need to change the complex background 3-form ${\cal
H}^0_{(3)}={\cal U}^r H^0_{(3)r}$ ("background" here meant in the sense
"$F=0$", not to be confused with asymptotic values).

We work in the gauge ${\rm Im}({\cal U}^2)=0$, corresponding to
\eq{{\cal U}^1=e^{\phi/2}\chi+i\,e^{-\phi/2}\,\,,
\quad{\cal U}^2=e^{\phi/2}\komma} which forces us to have a 
non-vanishing U(1) gauge field (see the comment at the end of the section).
Writing $ds^2=g_{\mu\nu}dx^\mu dx^\nu+g_{pq} dy^p dy^q$, the longitudinal 
metric $g_{\mu\nu}$ will get $F^2,F^4$ and $t_2, t_4, {\rm det}(F)$
corrections, whereas the transverse metric $g_{pq}\equiv
c^2\,\delta_{pq}$, just gets $t_2, t_4, {\rm det}(F)$ corrections,
since $F$ lies in the longitudinal directions.
It is convenient to work with the matrix $A$, defined from the
vielbein as $A\equiv {\rm log}\, e$ (where $e$ means $e_\mu^i$). It is
easy to get the metric from $A$. From the structure of the Einstein
equations we get $c= {\rm exp}(-\tr(A))$, and the Ricci tensor now
takes the simple form
\eqa{	R_{pq}&=-\*_p\D\*_q\D\left(\tr(A'{}^2)
	+\fraction{2}(\tr A')^2\right) +\fraction{2}\d_{pq}(\*\D)^2\tr
		A''\komma\cr
		R_{\mu\nu}&=-c^{-2}(\*\D)^2e_\mu{}^ie_\nu{}^jA''_{ij}\komma\cr
		}
where prime indicates differentiation with respect to $\Delta$.

The coordinate dependence of the solution can be written in terms of
$\Delta$, and the Ansatz thus takes the form
\eqa{
	&A=a_0\,\eta+a_2\,F^2+a_4\,F^4\komma\cr &{\cal
	H}_{(3)}=h\ast_y d\Delta-{\cal F}_2\wdg d\Delta\komma\cr
	&H_{(5)}=G_{(2)}\wdg \ast_y d\Delta-d\Delta\wdg\ast_x
	G_{(2)}\komma\cr &P=p\,d\Delta\,\qquad Q=q\,d\Delta\komma\cr}
where
\eqa{
	&({\cal
	F}_{(2)})_{ij}=f_1\,F_{ij}+f_3\,F^3_{ij}+f_5\,F^5_{ij}\komma\cr
	&(G_{(2)})_{ij}=g_1\,F_{ij}+g_3\,F^3_{ij}+g_5\,F^5_{ij}\cr} 
and we have chosen to write the 5-form in terms of the 2-form
\eq{G_{(2)}=-\ast G_{(4)}\punkt}
It is a straight-forward group-theoretical excercise to convince oneself
that no other independent combinations of $F$'s can enter the Ansatz.
$Q$ is pure gauge, since the form of $P$ yields $dQ=0$. The
Maurer--Cartan equations are therefore automatically satisfied.  Here
$p,q,h,f_1,f_3$ and $f_5$ are complex functions of $\Delta, t_2,t_4$
and ${\rm det}(F)$, and $a_0,a_2,a_4,g_1,g_3$ and $g_5$ are real
functions of the same variables and parameters. The functions $p$ and
$q$ should of course not be confused with the SL(2;$\Z$) charges
$(p,q)$.

Inserting the Ansatz in the equations of motion and Bianchi identities
yields equations for these functions, see appendix A. The background
solution corresponds to
\eq{a_0=-\fraction{8}\,{\rm log}\,\Delta\,,\quad p=\Fraction{i}{4}\,
\Delta^{-1}\,,\quad h=-\Fraction{i}{2}\,\Delta^{-1}\komma} 
and the rest zero. The first-order correction is the zero-mode
solution found in the previous section,
$f_1,g_1\sim\Delta^{-3/2}$, where the $\Delta$-dependence of the 
vielbeins has been taken into account. Now we can solve the equations order by
order and the result is expansions in negative powers of $\Delta$, see
appendix A.  Because of the complexity of the equations and the
Ansatz, we made a Mathematica program to get the expansion of the
solution. We were able to sum up these series and we have checked the
result explicitly for the $P$ equation of motion. The solution can be
written in terms of the following functions:
\eqa{
	&f_{\sss{2+}}\equiv \Delta+\Fraction{9}{16}\,t_2\komma\cr
	&f_{\sss{2-}}\equiv \Delta-\Fraction{9}{16\,}t_2\komma\cr
	&f_{\sss{4}}\equiv
	\Delta^2+(\Fraction{9}{16})^2\,\big((t_2)^2-4\,
	t_4\big)\komma\cr &f_{\sss{\rm det}}\equiv f_{\sss{2+}}
	f_{\sss{4}}-(\Fraction{27}{8})^2\, \hbox{\rm det}(F)\punkt\cr}
Only the first four terms in the expansions were used to
determine the solution. For half of the functions in the
Ansatz we have calculated 8-9 more terms and for the other
half 21-22 more terms, and they all agree with the solution
written in closed form. We therefore feel confident that the
solution is indeed the exact one.

We converted the expansions for $a_0,a_2$ and $a_4$ to expansions for
the metric with the following exact result
\eqa{
	&g_{\mu\nu}=(f_{\sss{\rm{det}}})^{-3/4}f_{\sss{4}}\,
	\eta_{\mu\nu}+\Fraction{9}{4}\,(f_{\sss{\rm{det}}})^{-3/4}
	f_{\sss{2-}} (F^2)_{\mu\nu}+(\Fraction{9}{4})^2
	(f_{\sss{\rm{det}}})^{-3/4}(F^4)_{\mu\nu}\komma\cr
	&c^2=(f_{\sss{\rm{det}}})^{1/4}\komma\cr} where
	$F_{\mu\nu}=\delta^i_\mu\delta^j_\nu\,F_{ij}$. For the 5-form, we 
get the remarkable result that $g_3$ and $g_5$ are identically zero, 
and therefore $G_{(2)}$ is proportional to $F$
\eqal
{(G_{(2)})_{ij}=\Fraction{3}{8}\,(f_{\sss{\rm{det}}})^{-1/2}F_{ij}\punkt}
{\thefiveform} 
The 3-form is specified by
\eqa{
	&h=-\fraction{2}\,(f_{\sss{4}})^{-1/2}\,\big(\Fraction{27}{8}\,
{\rm{pf}}(F)\,(f_{\sss{\rm{det}}})^{-1/2}+i\,\big)\komma\cr
	&f_1=-\Fraction{3}{4}\,(f_{\sss{\rm{det}}})^{-1/2}
(f_{\sss{4}})^{-1/2}f_{\sss{2-}}+i\,\Fraction{2}{9\rm{pf}(F)}\,
\big((f_{\sss{\rm{det}}})^{-1}(f_{\sss{4}})^{3/2}-(f_{\sss{4}})^{-1/2}
f_{\sss{2-}}\big)\komma\cr
	&f_3=-\Fraction{27}{16}\,(f_{\sss{\rm{det}}})^{-1/2}
(f_{\sss{4}})^{-1/2}-i\,\fraction{2\rm{pf}(F)}\,(f_{\sss{4}})^{-1/2}
\big(1-(f_{\sss{\rm{det}}})^{-1}f_{\sss{4}}f_{\sss{2-}}\big)\komma\cr
	&f_5=i\,\Fraction{9}{8\rm{pf}(F)}\,(f_{\sss{\rm{det}}})^{-1}
(f_{\sss{4}})^{1/2}\komma\cr}
where we have introduced the Pfaffian, which satisfies 
${\rm pf}(F)^2={\rm det}(F)$.  The 1-forms are
\eqal{
	&q=\Fraction{27}{8}\,\Delta\,{\rm{pf}}(F)\,
(f_{\sss{\rm{det}}})^{-1/2}(f_{\sss{4}})^{-1}\komma\cr
	&p=i\,\Delta\big((f_{\sss{4}})^{-1}-\fraction{4}(2\Delta
	f_{\sss{2+}}+f_{\sss{4}})(f_{\sss{\rm{det}}})^{-1}\big)
+\Fraction{27}{8}\,\Delta\,{\rm{pf}}(F)\,(f_{\sss{\rm{det}}})^{-1/2}
(f_{\sss{4}})^{-1}\punkt\cr}{\PQsolution}
The above solution for the metric, the 5-form, the 3-form and
the 1-forms constitutes the SL(2;$\Z$)-covariant part of the solution and
is therefore valid for all $(p,q)$ 5-branes. The only remaining
information lies in the background value of the scalars.
For instance, the NS 5-brane with $\tau_\infty=i$ has
\eqa{
	&\chi=\Fraction{27}{8}\,{\rm{pf}}(F)\,(f_{\sss{4}})^{-1}\komma\cr
	&e^{-\phi}=(f_{\sss{\rm{det}}})^{1/2}(f_{\sss{4}})^{-1}\komma\cr}
which in our choice of gauge corresponds to the scalar doublet
\eqa{
	&{\cal
	U}^1=(f_{\sss{\rm{det}}})^{1/4}(f_{\sss{4}})^{-1/2}
\big(\Fraction{27}{8}\,{\rm{pf}}(F)\,(f_{\sss{\rm{det}}})^{-1/2}+i\,\big)
\komma\cr
	&{\cal U}^2=(f_{\sss{\rm{det}}})^{-1/4}(f_{\sss{4}})^{1/2}\punkt\cr}
It is straightforward to get the doublet of 3-forms from
${\cal H}_{(3)}$ and the scalar doublet. As explained in
section \prel, a $(p,q)$ 5-brane can be obtained by
simultaneously making SL(2;$\Z$) transformations on these two
doublets. Then we of course just get the 5-branes obtained as the 
SL(2;$\Z$) orbit on $\tau_\infty=i$. For the solution with general 
charges 
and arbitrary background scalars, the scalar doublet is the finite 
field 
strength modification of (\genbgsolution), leading to the covariant 
expression
\eql{{\cal 
U}^r=\epsilon^{rs}\,\Big(k^{-1}\tilde{p}_s\,(f_{\sss{\rm{det}}})^{1/4}
\,(f_{\sss{4}})^{-1/2}\big(\Fraction{27}{8}\,{\rm{pf}}(F)
\,(f_{\sss{\rm{det}}})^{-1/2}+i\,\big)-k 
p_s\,(f_{\sss{\rm{det}}})^{-1/4}
\,(f_{\sss{4}})^{1/2}\Big)\punkt}\generalscalars
With the covariant gauge choice discussed below, we thus have an 
SL(2;$\Z$)-covariant solution.

To linear order the exact solution reduces to the zero-mode solution
found in section \zmsol.  For all the fields, except the 3-form on the
brane, it can easily be checked that we obtain the background NS
5-brane solution by setting $F$ and the parameters depending on $F$ to
zero. For the 3-form on the brane we have to be more careful, since we
divide by the Pfaffian in the imaginary parts. If we choose a certain
basis for $F$, see the next section, and insert $f_1,f_3$ and $f_5$
into ${\cal F}_{(2)}$, it can be seen that ${\cal F}_{(2)}$ will
indeed vanish when $F$ does.

A comment on U(1) gauge choices: The solutions for vanishing field 
strength,
given in eqs. (\bgsolution) and (\genbgsolution), 
use $Q=0$ as a choice of gauge for the local
U(1) symmetry. This gauge choice has the advantage of being manifestly
SL(2;$\Z$)-invariant. In the case of an NS5-brane, it is equivalent
to Im$({\cal U}^2)=0$. The reason for our departure from $Q=0$ when 
the tensor
is turned on is purely practical. The field configuration is always 
such
that $P$ points in the radial direction, which means that $Q$ is 
closed,
$dQ=0$, and it is formally trivial to perform a gauge transformation 
to
reach $Q=0$. The difficulty we encounter for finite field strength is
that even though we have an explicit expression for $Q$, 
eq. (\PQsolution), we have
not been able to integrate it to a closed analytic expression. The 
gauge
choice Im$({\cal U}^2)=0$ is therefore better suited to our ambition 
of
giving closed explicit solutions for all fields. It is not 
SL(2;$\Z$)-covariant as it stands, but can easily be made so 
with the introduction
of general charges---it then reads Im$({\cal U}^r\tilde p_r)=0$.
The general solution (\generalscalars) clearly fulfills this
equation.

\section\metric{Properties of the metric}In this section we analyse 
the 
metric arising from the deformation of the 5-brane via the finite 
constant 
field strength. To get a solution which makes sense, we must impose a 
condition on the parameters of this field strength.

Since $F$ is antisymmetric, we can find a basis where it takes the
simple form
\eq{F_{ij}=\mx{0&\nu_1&0&0&0&0\\-\nu_1&0&0&0&0&0\\0&0&0&\nu_2&0&0\\0&0&
-\nu_2&0&0&0\\0&0&0&0&0&\nu_3\\0&0&0&0&-\nu_3&0\\}\punkt}
This is defined in Minkowski space and therefore we get a sign
change in $t_2$ compared to the trace in Euclidean space
\eqa{
	&t_2=2 \nu_1^2-2 \nu_2^2-2 \nu_3^2\,,\qquad t_4=2 \nu_1^4+2
	\nu_2^4+2 \nu_3^4\komma\cr &{\rm
	pf}(F)=-\nu_1\,\nu_2\,\nu_3\,,\qquad\qquad\, {\rm
	det}(F)=\nu_1^2\, \nu_2^2\,\nu_3^2\punkt\cr} 
In this basis, the
metric in the brane directions will be diagonal, since $F^2$
and $F^4$ are diagonal. Furthermore, from the structure of
$F$, we get a natural split into three two-dimensional spaces,
one Minkowskian and two Euclidean. The components of the
metric in the subspaces are
\eqa{
	&g_{11}=(f_{\sss {\rm
	det}})^{-3/4}\Big(\Delta(\Delta+\Fraction{9}{4}\nu_1^2)
+(\Fraction{9}{8})^2\big(\nu_1^4-(\nu_2^2-\nu_3^2)^2\big)\Big)\komma\cr
	&g_{33}=(f_{\sss {\rm
	det}})^{-3/4}\Big(\Delta(\Delta-\Fraction{9}{4}\nu_2^2)
+(\Fraction{9}{8})^2\big(\nu_2^4-(\nu_1^2+\nu_3^2)^2\big)\Big)\komma\cr
	&g_{55}=(f_{\sss {\rm
	det}})^{-3/4}\Big(\Delta(\Delta-\Fraction{9}{4}\nu_3^2)
+(\Fraction{9}{8})^2\big(\nu_3^4-(\nu_1^2+\nu_2^2)^2\big)\Big)\komma\cr}
where $f_{\sss {\rm det}}$, expressed in the $\nu$ parameters,
becomes
\eq{f_{\sss {\rm 
det}}=\big(\Delta+\Fraction{9}{8}(\nu_1^2-\nu_2^2-\nu_3^2)
\big)\big(\Delta^2-\Fraction{81}{64}(\nu_1^4+(\nu_2^2-\nu_3^2)^2
+2\nu_1^2(\nu_2^2+\nu_3^2))\big)-(\Fraction{27}{8})^2\nu_1^2\, 
\nu_2^2\,\nu_3^2\punkt}

Besides from the advantage of having a diagonal metric, one also has
detailed knowledge about $F$. When all the $\nu$'s are non-vanishing, 
the matrix has full
rank, and lower rank can be obtained by turning one or two of the
parameters off. The exact solution found in the previous section is of
course simplified when using a lower rank matrix. If we for instance
set $\nu_2$ and $\nu_3$ to zero, everything can be expressed in terms
of just two functions $\Delta\pm\Fraction{9}{8}\nu_1$.

In the limit of vanishing 2-form field strength, the longitudinal
metric has the usual asymptotic Minkowski form $(-,+,+,+,+,+)$. For
finite field strength, the metric should still asymptotically be 
Minkowski, \ie, the first component should be negative and the rest
should be positive. This gives a restriction on the $\nu$
parameters. Because of the negative power of $f_{\sss {\rm det}}$ in
the metric, in order to get a non-singular solution, the parameters
should fulfill yet another restriction. This is related to the
breakdown of Born--Infeld dynamics at some finite field strength. It
turns out that the latter requirement is the stronger, and we get
\eq{\nu_1^2+\nu_2^2+\nu_3^2\leq\Fraction89\punkt}
In ref. [\Finite], we calculated the finite tensor deformations of 
the D3
and the M5-branes. Because of the duality properties, the solutions
just depend on one parameter $\nu$, which also has an upper bound. In
the limiting case, the metric blows up asymptotically in some
directions on the brane and shrinks in the other directions. This was
interpreted as smeared out strings or membranes. Here we get the same
kind of behaviour in the limiting case $\nu_1^2+\nu_2^2+\nu_3^2=
{8\over9}$. For almost every possibility (the exception is considered 
below), after trivial rescalings, the leading asymptotic terms in the 
metric become
\eq{ds^2=\rho^{3/2}dx_1^2+\rho^{-1/2}(dx_2^2+dx_3^2+dy^2)\komma}
where $dx_i^2$, $i=1,2,3$, corresponds to the three 
two-dimensional subspaces on the
brane. We see that the metric in the two-dimensional Minkowskian
subspace blows up asymptotically, whereas it shrinks in the other
directions, and the result is a smeared out string. The exception is 
when just $\nu_2$ or $\nu_3$ is different from zero. Considering the 
latter, we get the following asymptotical behaviour of the metric
\eq{ds^2=\rho\,(dx_1^2+dx_2^2)+\rho^{-1}(dx_3^2+dy^2)\punkt}
Here four directions blow up and the rest shrink. This is 
interpreted as a smeared out 3-brane.

In summary, we saw in the previous section that $F$ and its higher
powers were used as a basis, and the solution was described in terms 
of the
traces and the determinant. Using the frame introduced in this 
section,
various properties of the solution become more clear. The metric
becomes diagonal and we have seen that we get an upper bound on the
parameters of the finite constant field strength.

\section\disc{Discussion}We have constructed explicitly the type \II B
supergravity solutions corresponding to 5-branes with arbitrary
charges 
and with finite Born--Infeld field strength, or equivalently, in
a finite background tensor field. The solutions are considerably more
involved than in the previously treated cases of type \II B 3-branes
and M5-branes [\Finite], due to the fact that a $(p,q)$ 5-brane
couples both to the NS-NS and RR 2-forms and to the chiral 4-form,
which in the solutions manifests itself as excitations of all these
tensors.

One of the main results of the paper is that it answers the question
provoked by the title---what is ``the $B$-field'' that a $(p,q)$ 
5-brane feels? We have parametrised the physical modes by a real 
2-form, which identifies the unique combination of the doublet of 
2-forms and the 4-form that can not be gauged away in the presence
of a $(p,q)$ 5-brane.

We envisage some potential applications of the result.  The first one,
which was one of the main reasons for initiating this work, concerns
the SL(2;$\Z$)-covariant formulation of type \II B 5-brane dynamics,
continuing the work of refs. [\CT,\CW] 
for lower-dimensional type \II B branes. 
Related issues have been addressed in a number of papers
[\DualActions,\CW,\LuRoy,\WeWy].  A straight-forward Poincar\'e
dualisation of the Born--Infeld action was first attempted in
ref. [\DualActions].  It is however unclear how relevant such a
procedure is---the Born--Infeld field strength, being related to the
NS-NS 2-form, is not SL(2;$\Z$)-invariant and the dual can hardly be
associated with the SL(2;$\Z$)-invariant 4-form. In ref. [\CW], it was
clarified how a duality relation between a complex 2-form and a real 
4-form
will be a natural ingredient in the type of description we are looking
for.  For all values of the radius, there will be a (radius-dependent)
duality relation. It is likely that this relation may be deduced using
the results of the present paper. The only obstacle could be that one
encounters equations that do not have obvious analytic solutions, as
in ref. [\DualActions].  We find it more likely, however, that this
will happen when one tries to solve for some tensors in terms of 
others. 
The attempt of ref. [\DualActions] to dualise the Born--Infeld field
failed due to the occurrence of quintic equations.
Here, we also note that both the 2-form and 4-form field strengths on
the brane are explicitly parametrised by a real 2-form ($F$),
which, in light of the surprisingly simple expression 
(\thefiveform) for the 4-form
(for which we have no other explanation than that we were lucky in
our choice of basis), means that the complex 2-form ${\cal F}$ is 
easily
expressed in terms of the 4-form $G$. The opposite, to express $G$ in
terms of ${\cal F}$ would demand solving a quintic equation.
An order-by-order solution of the problem of finding an 
SL(2;$\Z$)-covariant action was attempted in
ref. [\WeWy], with partial success.

It is well known that the Born--Infeld action (and similar actions,
\eg\ for chiral 2-forms in $D=6$) has a maximal finite field
strength where the generalisation of the volume element containing
$\det(g+F)$ degenerates, and the canonical electric field
diverges. Analogously, there is a maximal value of the
non-commutativity parameter in non-commutative Maxwell theory.  The
solutions presented here and in ref. [\Finite] go further---they
investigate in which sense 10- or 11-dimensional space-time
degenerates in this limit.  The result is, maybe somewhat
surprisingly, that the limiting space-times are well-defined, but with
an asymptotic structure that differs radically from the asymptotically
Minkowski brane solutions. There is an effective reduction of the
dimensionality of the boundary (boundary understood in the same sense
as in AdS space), corresponding to a lower-dimensional brane. In the
present paper string and 3-brane degeneracies were presented 
explicitly, 
which gives an even stronger indication that the limiting cases in 
some
not yet fully understood sense correspond to ``branes-within-branes''
situations.  
It is conceivable that there is
an AdS/CFT-like correspondence between superstring theory on these
spaces and field theories on the (codimension $>1$) boundaries.

Finally, the solutions found here and in ref. [\Finite] should be
relevant in connection to non-commutative
geometry [\SeibergWittenNC]. It has already been stressed
that a finite value of the brane field strength is identical to a
situation with finite background $B$ field (due to the gauge
transformation $\delta B=d\Lambda$, $\delta A=\Lambda$, leaving the
combination $dA-B$ invariant). 
The results of this paper contain explicit information, presented in
an SL(2;$\Z$)-covariant form, about where
among the various tensor fields in type \II B supergravity the
non-commutativity parameter is to be found. The corresponding 
information for the 3-brane is found in ref. [\Finite]. Although the
full solution for the M5-brane is known, we do not know the 
generalisation of non-commutativity relevant for (chiral) 2-forms,
and in that sense the M5-brane is still concealing some, probably
very interesting and stringy, secrets.
To our knowledge, non-commutative field
theory has never been treated with the ambition of manifesting
SL(2;$\Z$)---this may prove interesting for \eg\ the 3-brane and
5-branes in type \II B, and the present techniques would undoubtedly 
be
relevant. Neither has any concrete understanding been obtained 
concerning generalisations to higher rank tensors.
Maybe detailed knowledge of the target space configurations
of the type presented here and in ref. [\Finite] can be helpful.

\appendix{Equations and Expansions}In this appendix we write down the 
equations obtained by insertion of the finite tensor Ansatz and the 
resulting solution, obtained using Mathematica, 
as expansions in negative powers of $\Delta$. 
The equations are
\eqa{
	&p'-2i\,q p-i\,h^2+\Fraction{i}{2}\,\tr({\cal
	F}_{(2)}^2)=0\komma\cr &{\cal F}_{(2)}'-2\,A'{\cal
	F}_{(2)}+i\,q{\cal F}_2-i\,p\bar{{\cal
	F}}_{(2)}-4i\,hG_{(2)}-4i\,G_{(2)}\star{\cal
	F}_{(2)}=0\komma\cr &h'-h\tr(A')-i\,q h+i\bar{h}p=0\komma\cr
	&2\,G_{(2)}A'+G_{(2)}'-\tr(A')G_{(2)}+{\rm Im}(h{\cal
	F}_{(2)})=0\komma\cr
	&\tr\big((A')^2\big)+\fraction{2}\,\big(\tr(A')\big)^2+2\,\vert
	p\vert^2-\tr(\bar{{\cal F}}_{(2)}{\cal
	F}_{(2)})+4\,\tr(G_{(2)}^2)-2\,\vert h\vert^2=0\komma\cr
	&A''-2\,\bar{{\cal F}}_{(2)}{\cal
	F}_{(2)}+\fraction{4}\,\tr(\bar{{\cal F}}_{(2)}{\cal
	F}_{(2)})\,\eta-\fraction{2}\,\vert
	h\vert^2\,\eta+2\,\tr(G_{(2)}^2)\,\eta-8\,G_{(2)}^2=0\komma\cr}
where $(G_{(2)}\star{\cal
F}_{(2)})_{ij}=\fraction{4}\,\epsilon_{ijklmn}G_{(2)}{}^{kl}{\cal
F}_{(2)}{}^{mn}$, and prime means differentiation with respect
to $\Delta$. The first two equations are the equations of
motion for $P$ and ${\cal H}_{(3)}$, respectively. The next
two are the Bianchi identities for ${\cal H}_{(3)}$ and
$H_{(5)}$, and the last two are the Einstein equations. In
fact, there is a redundancy in these equations, since \eg\ the
first Einstein equation differentiated yields a combination of
the others.

The above equations are written on a compact form, since we really
should split them according to the basis elements $F, F^3$ and $F^5$
for the antisymmetric matrices and $\eta, F^2$ and $F^4$ for the
symmetric matrices. The first few terms in the solution are
\eqa{
	&a_0=-\fraction{8}\,{\rm 
log}\,\Delta-\Fraction{3}{8}\Fraction{9}{16}\,
t_2\,\Delta^{-1}+\fraction{16}(\Fraction{9}{16})^2\,(5
	t_2^2-8t_4)\,\Delta^{-2}+\cdots\komma\cr
	&a_2=\Fraction{9}{8}\,\Delta^{-1}-\Fraction{9}{8}\Fraction{9}{16}\,
t_2\,\Delta^{-2}+\Fraction{3}{8}(\Fraction{9}{16})^2\,(t_2^2+4t_4)\,
\Delta^{-3}+\cdots\komma\cr
	&a_4=(\Fraction{9}{8})^2\,\Delta^{-2}
-\Fraction{27}{32}\Fraction{9}{16}\,t_2\,\Delta^{-3}
+\Fraction{9}{32}(\Fraction{9}{16})^2\,t_4\,\Delta^{-4}+\cdots\komma\cr
	&f_1=-\Fraction{3}{4}\,\Delta^{-3/2}
\big(1+\Fraction{3}{2}\Fraction{9}{16}\,t_2\,\Delta^{-1}+\cdots\big)
+i\,\fraction{16}\Fraction{9}{16\,{\rm
	pf}(F)}\,(3t_2^2-4t_4)\,\Delta^{-2}+\cdots\komma\cr
	&f_3=i\,\fraction{{\rm
	pf}(F)}\Delta^{-2}\big(-\Fraction{9}{16}\,t_2
+(\Fraction{9}{16})^2\,t_2^2\,\Delta^{-1}+\cdots\big)
-\Fraction{27}{16}\,\Delta^{-5/2}+\cdots\komma\cr
	&f_5=i\,\Fraction{9}{8\,{\rm
	pf}(F)}\Delta^{-2}\big(1-\Fraction{9}{16}\,t_2\,
\Delta^{-1}+\fraction{2}(\Fraction{9}{16})^2\,(t_2^2+4t_4)\,
\Delta^{-2}+\cdots\big)\komma\cr
	&g_1=\Fraction{3}{8}\,\Delta^{-3/2}
\big(1-\Fraction{9}{16}\,t_2\,\Delta^{-1}-\Fraction{3}{8}(\Fraction{9}{16})^2\,
(t_2^2-16t_4)\,\Delta^{-2}+\cdots\big)\komma\cr
	&g_3=g_5=0+\cdots\komma\cr
	&p=i\,\Fraction{1}{4}\,\Delta^{-1}
\big(1+\Fraction{9}{16}\,t_2\,\Delta^{-1}+\cdots\big)+\Fraction{27}{8}\,{\rm
	pf}(F)\,\Delta^{-5/2}+\cdots\komma\cr
	&q=\Fraction{27}{8}\,{\rm
	pf}(F)\,\Delta^{-5/2}\big(1-\fraction{2}\Fraction{9}{16}\,t_2\,{~\rm
	pf}(F)\,\Delta^{-1}+\cdots\big)\punkt\cr} 
From the expansion
of $p$ and $q$ we get an expansion of the scalar doublet
${\cal U}^r$, using the constraint and the gauge choice. The
result is, for charges $(1,0)$ and $\tau_\infty=i$,
\eqa{
	&{\cal
	U}^1=i\,\Delta^{-1/4}+i\,\fraction{4}\Fraction{9}{16}\,t_2\,
\Delta^{-5/4}+\Fraction{27}{8}\,{\rm
	pf}(F)\,\Delta^{-7/4}+\cdots\komma\cr &{\cal
	U}^2=\Delta^{1/4}-\fraction{4}\Fraction{9}{16}\,t_2\,
\Delta^{-3/4}+\fraction{32}(\Fraction{9}{16})^2\,(13t_2^2-32t_4)\,
\Delta^{-7/4}+\cdots\punkt\cr}
By calculating the charges, it can be checked that we still
have a (1,0)-brane with $\tau_\infty=i$.
Covariantisation is straight-forward.

\vfill\eject
\refout

\end